\documentclass{svjour2}                    
\smartqed  
\usepackage{amsmath}
\usepackage{graphicx}
\usepackage{dsfont}
\usepackage[colorlinks=true,breaklinks=true,allcolors=blue]{hyperref}
\usepackage{doi}
\usepackage{mathptmx}      
\usepackage{bm}
\newcommand\one{{\mathds{1}}}

\newcommand\RR{{\mathds{R}}}
\newcommand\CC{{\mathds{C}}}
\newcommand\rmi{{\mathrm{i}}}
\newcommand\rme{{\mathrm{e}}}
\newcommand\rmd{{\mathrm{d}}}

\newcommand{\arctanh}{\mathrm{\,arctanh\,}}

\newcommand{\bsigma}{\bm\sigma}

\DeclareMathOperator{\Tr}{{Tr}}
\begin{document}

\title{Microcanonical analysis of the Curie-Weiss anisotropic quantum Heisenberg model in a magnetic field
}

\titlerunning{Microcanonical Curie-Weiss quantum Heisenberg model in a magnetic field}       

\author{Gerrit Olivier         \and
        Michael Kastner 
}


\institute{G. Olivier \at
              ISTerre, Universit\'e Joseph Fourier, Grenoble, France,
              and Institute of Mine Seismology, Stellenbosch 7600, South Africa.
           \and
           M. Kastner \at
              National Institute for Theoretical Physics (NITheP), Stellenbosch 7600, South Africa,
              and Institute of Theoretical Physics,  University of Stellenbosch, Stellenbosch 7600, South Africa.
              \email{kastner@sun.ac.za}
}

\date{Received: date / Accepted: date}

\maketitle

\begin{abstract}
The anisotropic quantum Heisenberg model with Curie-Weiss-type interactions is studied analytically in several variants of the microcanonical ensemble. (Non)equivalence of microcanonical and canonical ensembles is investigated by studying the concavity properties of entropies. The microcanonical entropy $s(e,\bm{m})$ is obtained as a function of the energy $e$ and the magnetization vector $\bm m$ in the thermodynamic limit. Since, for this model, $e$ is uniquely determined by $\bm m$, the same information can be encoded either in $s(\bm{m})$ or $s(e,m_1,m_2)$. Although these two entropies correspond to the same physical setting of fixed $e$ and $\bm m$, their concavity properties differ. The entropy $s_{\bm h}(u)$, describing the model at fixed total energy $u$ and in a homogeneous external magnetic field $\bm h$ of arbitrary direction, is obtained by reduction from the nonconcave entropy $s(e,m_1,m_2)$. In doing so, concavity, and therefore equivalence of ensembles, is restored. $s_{\bm h}(u)$ has nonanalyticities on surfaces of co-dimension 1 in the $(u,\bm{h})$-space. Projecting these surfaces into lower-dimensional phase diagrams, we observe that the resulting phase transition lines are situated in the positive-temperature region for some parameter values, and in the negative-temperature region for others. In the canonical setting of a system coupled to a heat bath of positive temperatures, the nonanalyticities in the microcanonical negative-temperature region cannot be observed, and this leads to a situation of effective nonequivalence even when formal equivalence holds.
\end{abstract}

\section{Introduction}

In this paper we report a microcanonical study of the anisotropic quantum Heisenberg model with Curie-Weiss-type interactions in the presence of an external magnetic field. The study extends and generalizes the results reported in an earlier work \cite{KastnerJSTAT10}. The model and the setting are motivated by recent and envisaged experiments with trapped ions, or with ultracold atoms or molecules in optical lattices. Dipolar gases in optical traps have been suggested as laboratory realizations of lattice spin models where the coupling parameters can be tuned, allowing for the realization of Hamiltonians which are of interest in condensed matter physics \cite{Micheli_etal06}. Recently, trapped ions have been used to engineer one- and two-dimensional lattices of long-range interacting spins \cite{Friedenauer_etal08,Islam_etal11,Lanyon_etal11,Britton_etal12}. However, as distinguished from their condensed matter counterparts, ultracold gases are extremely well isolated from their environment. Accordingly, an appropriate description of their equilibrium properties should be within the microcanonical ensemble \cite{GrossHolt96,Kastner10}.

For systems with short-range interactions, the choice of the statistical ensemble is typically of minor importance and can be considered a finite-size effect: differences between, say, microcanonical and canonical quantities are known to vanish in the thermodynamic limit of large system size, and the various statistical ensembles become equivalent \cite{Ruelle}. In the presence of long-range interactions this is in general not the case, and microcanonical and canonical approaches can lead to different thermodynamic properties even in the infinite-system limit \cite{TouElTur04}. Here, long-range refers to interactions decaying asymptotically like $r^{-\alpha}$ for large distances $r$, where the exponent $\alpha$ satisfies $0\leq\alpha\leq d$ and $d$ is the spatial dimension of the system. In the astrophysical context where long-range interactions prevail, nonequivalence of ensembles and the importance of microcanonical calculations have long been known for gravitational systems \cite{LynWood68,Thirring70}. Nonequivalence of ensembles is usually accompanied by unfamiliar thermodynamical properties in the microcanonical ensemble. An example is the occurrence of negative microcanonical specific heat, indicating that---quite counterintuitively---temperature will decrease when energy is pumped into the system.

Nonequivalence of ensembles has been studied almost exclusively in the classical mechanical context. A notable example is the paper by Pflug \cite{Pflug80} on gravitating fermions, where a negative specific heat is found for all negative values of the energy. Inspired by experimental efforts to emulate long-range interacting quantum spin systems by means of ultracold gases, the aim of the present paper is to contribute to the understanding of nonequivalent ensembles in quantum spin systems. In experimental realizations, the long-range exponent $\alpha$ is equal to 3 for dipole--dipole interactions, but it can be as low as $0.05$ for trapped ion-based quantum simulators of long-range Ising models \cite{Britton_etal12}. It is well-known that many equilibrium properties of long-range interacting systems with small values of $\alpha$ are well modeled by Curie-Weiss-type interactions, i.e.\ long-range interactions with $\alpha=0$, and such a choice renders the model analytically solvable \cite{PearceThompson75,GiansantiMoroniCampa02}. Moreover, the canonical free energy for a large class of long-range interacting models has been shown to coincide with the Curie-Weiss results in the thermodynamic limit \cite{Mori11,Mori12}.

In this paper we consider the anisotropic quantum Heisenberg model with Curie-Weiss-type interactions, which is a model of anisotropically interacting spin-$1/2$ degrees of freedom. In reference \cite{KastnerJSTAT10}, the microcanonical entropy $s(e,m_3)$ was computed in the thermodynamic limit as a function of the energy $e$ and the $3$-component $m_3$ of the magnetization.\footnote{A referee of the present paper pointed out several nonrigorous steps in \cite{KastnerJSTAT10}. We will comment on this in Sec.\ \ref{s:sem}.} Depending on the values of the anisotropy parameters in the Hamiltonian, this function was found to be concave in some cases, and nonconcave in others. The latter is a hallmark of nonequivalent ensembles, implying that some of the equilibrium states of the (generalized) microcanonical ensemble at fixed $e$ and $m_3$ cannot be observed as canonical equilibrium states at any temperature $T$ and external magnetic field $\bm{h}=h\bm{e}_3$ in the $3$-direction \cite{Touchette11}. However, in realistic physical situations the magnetization component $m_3$ is usually not conserved. This observation motivated the present work, namely the study of an ensemble with fixed energy and magnetic field, but fluctuating magnetization.

Besides this motivation from the experimental side, the results reported in this paper contribute several novel aspects, and reveal several pitfalls, related to nonequivalent statistical ensembles. For the Curie-Weiss anisotropic quantum Heisenberg model, $e$ is uniquely determined by $\bm m$. While both entropy functions, $s(\bm{m})$ and $s(e,m_1,m_2)$, correspond to the same physical setting of fixed $e$ and $\bm m$, the former is a concave function, while the latter is not. From the nonconcave entropy $s(e,m_1,m_2)$, a concave $s_{\bm h}(u)$ is obtained by reduction, and equivalence of ensembles is restored. Depending on the values of the anisotropy parameters, $s_{\bm h}(u)$ shows a continuous phase transition either in the positive-temperature region, or in the negative-temperature region. In the canonical setting of a system coupled to a heat bath of positive temperatures, the nonanalyticities in the microcanonical negative-temperature region cannot be observed, and this leads to a situation of effective nonequivalence even when formal equivalence holds. It is remarkable to find in a single model such a variety of equivalence and nonequivalence situations, some of which had not been discussed in this context before.

\section{Curie-Weiss anisotropic quantum Heisenberg model}
\label{s:model}

The model that we study consists of $N$ spin-$1/2$ degrees of freedom, each of which is interacting with every other at equal strength (Curie-Weiss-type interactions). The corresponding Hilbert space $\mathcal{H}=(\CC^2)^{\otimes N}$ is the tensor product of $N$ copies of the spin-$1/2$ Hilbert space $\CC^2$, and the Hamiltonian operator is given by
\begin{equation}\label{e:Hamiltonian}
H_h=-\frac{1}{2N}\sum_{k,l=1}^N \left(\lambda_1 \sigma_k^1 \sigma_l^1 + \lambda_2 \sigma_k^2 \sigma_l^2 + \lambda_3 \sigma_k^3 \sigma_l^3\right) - \bm{h}\cdot\sum_{k=1}^N \bsigma_k.
\end{equation}
The $\sigma_k^\alpha$ are operators on $\mathcal{H}$ and act like the $\alpha$-component of the Pauli spin-$1/2$ operator on the $k$th factor of the tensor product space $\mathcal{H}$, and like identity operators $\one_2$ on all the other factors,
\begin{equation}
\sigma_k^\alpha=\one_2\otimes\cdots\otimes\one_2\otimes\underbrace{\sigma^\alpha}_{\hspace{-5mm}\mbox{$k$th factor}\hspace{-5mm}}\otimes\one_2\otimes\cdots\otimes\one_2,\qquad\alpha\in\{1,2,3\}.
\end{equation}
The resulting commutation relation is
\begin{equation}\label{e:commutator}
\bigl[\sigma_k^\alpha,\sigma_l^\beta\bigr]=2\rmi\, \delta_{k,l}\,\epsilon_{\alpha\beta\gamma}\sigma_k^\gamma,\qquad\alpha,\beta,\gamma\in\{1,2,3\},
\end{equation}
where $\delta$ denotes Kronecker's symbol and $\epsilon$ is the Levi-Civita symbol. The parameter $\bm h$ in the Hamiltonian is the magnetic field vector, and the constants $\lambda_1$, $\lambda_2$, and $\lambda_3$ determine the coupling strengths in the various spatial directions and allow to adjust the degree of anisotropy. It is convenient to introduce a collective spin operator $\bm{S}$ with components
\begin{equation}\label{e:Sa}
S_\alpha=\frac{1}{2}\sum_{i=1}^N \sigma_i^\alpha,\qquad\alpha\in\{1,2,3\},
\end{equation}
which allows us to rewrite the Hamiltonian \eqref{e:Hamiltonian} in the form
\begin{equation}\label{e:Hamiltoniancollective}
H_{\bm{h}}=-\frac{2}{N}\left(\lambda_1 S_1^2 + \lambda_2 S_2^2 + \lambda_3 S_3^2\right) - 2\bm{h}\cdot\bm{S}.
\end{equation}
The Hamiltonian \eqref{e:Hamiltonian} or \eqref{e:Hamiltoniancollective} differs from the one discussed in \cite{KastnerJSTAT10} by the fact that the magnetic field vector $\bm h$ is not necessarily along the $z$-direction, but can have any orientation in $\RR^3$.

Here we consider the coupling constants $\lambda_1$, $\lambda_2$, and $\lambda_3$ to be nonnegative, but otherwise arbitrary. The exact expression for the canonical Gibbs free energy $g$ as a function of the inverse temperature $\beta=1/T$ and the magnetic field $h$ is known for this model and is reported for example in \cite{PearceThompson75}.\footnote{Here and in the following Boltzmann's constant is set to unity.} The model is found to display a transition from a ferromagnetic to a paramagnetic phase in the canonical ensemble.

\section{Microcanonical entropy \texorpdfstring{$s(e,\bm{m})$}{s(e,m)}}
\label{s:sem}
Owing to the long-range character of the interactions in the Hamiltonian \eqref{e:Hamiltonian}, microcanonical and canonical ensembles do not necessarily yield equivalent results. This also implies that in general the microcanonical entropy cannot be obtained from the canonical free energy by Legendre transform, but has to be computed by other methods \cite{Touchette10}. Compared to the canonical case, such a calculation is known to be usually more difficult, as was already observed by Gibbs in his classical treatise \cite{Gibbs}.

Our aim is to compute the microcanonical entropy
\begin{equation}\label{e:sem3def}
s(e,\bm{m})=\lim_{N\to\infty}\frac{1}{N}\ln\Omega_N(e,\bm{m})
\end{equation}
as a function of the energy (per spin) $e$ and the vector of magnetization (per spin) $\bm{m}$, where $\Omega_N$ denotes the density of states (or microcanonical partition function) of the $N$-spin system. In general, the two operators $H$ and $\bm{S}$ corresponding to the thermodynamic variables $e$ and $\bm{m}$ do not commute, and in this case there is no consensus on the correct definition of $\Omega_N(e,\bm{m})$. Several proposals, with their advantages and drawbacks, have been discussed in Sec.\ 3 of reference \cite{KastnerJSTAT10}. Here we take the canonical partition function
\begin{equation}
Z_N(\beta,-\beta h)=\Tr\rme^{-\beta H_{\bm{h}}}
\end{equation}
as a starting point and define the density of states as its inverse Laplace transformation,
\begin{multline}\label{e:OmegaDef}
\Omega_N(e,\bm{m})=\frac{1}{(2 \pi \rmi)^4} \int_{a-\rmi\infty}^{a+\rmi\infty} \rmd(N \beta) \rme^{N \beta e}\\
\times \left(\prod_{\alpha=1}^3 \int_{b_\alpha-\rmi\infty}^{b_\alpha+\rmi\infty} \rmd(-N\beta h_\alpha)\right)\rme^{-N\beta \bm{h} \cdot \bm{m}} \Tr\rme^{-\beta H_{\bm{h}}}.
\end{multline}
The real constants $a$ and $b_\alpha$ have to be chosen such that, for each of the integrations, all poles of the integrand lie to the left of the integration contour in the complex plane, but otherwise the constants are arbitrary. This definition of $\Omega_N$ is not identical with the definition used in \cite{KastnerJSTAT10}, but it yields the same result for the entropy \eqref{e:sem3def} in the thermodynamic limit. While the classical density of states has the straightforward interpretation of counting the number of microstates with a given constraint, no such obvious interpretation seems to exist for our definition. The virtue of  definition \eqref{e:OmegaDef}, however, is that it preserves the familiar relation between microcanonical and canonical partition functions, the latter being the Laplace transform of the former for all finite system sizes $N$. Note that this definition via an inverse Laplace transform is related to, but not identical to computing the microcanonical entropy via Legendre-Fenchel transform from the canonical free energy in the thermodynamic limit. Whereas the latter fails when ensembles are nonequivalent, our definition \eqref{e:OmegaDef} does not suffer from this shortcoming.

\subsection{Evaluation of the trace}
\label{s:trace}
To further analyze the density of states, it is convenient to switch to real integrations by means of the substitutions $k=\rmi(a-\beta)$ and $\bm{l}=\rmi(\bm{b}+\beta\bm{h})$, yielding
\begin{multline}\label{e:Omega1}
\Omega_N(e,\bm{m})=\frac{N^4}{(2 \pi)^4}\int\rmd k \int\rmd^3 l\, \exp\left[N(a+\rmi k)e+N(\bm{b}+\rmi\bm{l})\cdot\bm{m}\right]\\
\times\Tr \left\{\exp\left[\frac{2}{N}(a+\rmi k)\mathcal{S}^2-2(\bm{b}+\rmi \bm{l})\cdot\bm{S}\right]\right\}.
\end{multline}
Here the anisotropic collective spin operators
\begin{equation}
\mathcal{S}_\alpha=\sqrt{\lambda_\alpha} S_\alpha
\end{equation}
 and $\mathcal{S}^2=\mathcal{S}_1^2+\mathcal{S}_2^2+\mathcal{S}_3^2$ have been introduced to ease the notation. Unless specified otherwise, domains of integration always extend over $\RR$ (for one-dimensional integrals) or $\RR^3$ (for volume integrals).

To evaluate \eqref{e:Omega1}, we rewrite the density of states \eqref{e:Omega1} in such a way that the $N$-spin trace decouples into a product of one-spin traces. The necessary steps are an adaptation to our microcanonical setting of techniques that have been used by Tindemans and Capel \cite{TinCa74} in a canonical calculation. These manipulations are analogous to those in Sec.\ 3.2.1 of \cite{KastnerJSTAT10} and are therefore not reported in detail here.%
\footnote{The derivation in Sec.\ 3.2.1 of \cite{KastnerJSTAT10} includes a step that is not rigorously justified, namely the rewriting of a product of exponentials of operators as an exponential of a sum of operators in Eq.\ (38) of \cite{KastnerJSTAT10}. While these two expressions clearly are not equal, it has been proved in Appendix A of \cite{TinCa74} that, in a similar situation, the neglected terms do not contribute in the thermodynamic limit to the integral under investigation. While we were not able to adapt this proof to our calculation, it appears plausible that a similar reasoning should also yield the correct result in a microcanonical calculation.}
Similar to equations (42) and (43) of \cite{KastnerJSTAT10}, the result is a $(3n+4)$-dimensional Laplace integral
\begin{multline}\label{e:Omega3}
\Omega_N(e,\bm m) = \frac{2^N N^4}{(2\pi)^4}\int_{a-\rmi\infty}^{a+\rmi\infty}\rmd s \left(\prod_{\alpha=1}^3\int_{b_\alpha-\rmi\infty}^{b_\alpha+\rmi\infty}\rmd t_\alpha\right)\\
\times\lim_{n\to\infty}\left(\frac{N}{2\pi ns}\right)^{3n/2}\int\!\cdots\!\int\rmd^3 x^{(1)}\cdots\rmd^3 x^{(n)} \exp\left[N\mathcal{F}\bigl(s,\bm{t},\bigl\{\bm{x}^{(i)}\bigr\}\bigr)\right]
\end{multline}
in the limit $n\to\infty$, but with exponent
\begin{equation}\label{e:F}
\mathcal{F}(s,\bm{t},\{\bm{x}^{(i)}\})=es+\bm{m}\cdot\bm{t}-\frac{1}{2ns}\sum_{i=1}^n \bm{x}^{(i)}\cdot\bm{x}^{(i)}+\ln\cosh \left[r\bigl(\bm{t},\bigl\{\bm{x}^{(i)}\bigr\}\bigr)\right],
\end{equation}
where
\begin{equation}
r\equiv r\bigl(\bm{t},\bigl\{\bm{x}^{(i)}\bigr\}\bigr) = \sqrt{c_1^2+c_2^2+c_3^2}
\end{equation}
and
\begin{equation}\label{e:c_alpha}
c_\alpha\equiv c_\alpha\bigl(t_\alpha,\bigl\{x_\alpha^{(i)}\bigr\}\bigr)=\frac{1}{n}\sqrt{\lambda_\alpha}\sum_{i=1}^n x_\alpha^{(i)} - t_\alpha.
\end{equation}
As expected, these expressions are similar to the corresponding ones in \cite{KastnerJSTAT10}. The most noteworthy difference is that \eqref{e:c_alpha} is more symmetric than its counterpart equation (39) in \cite{KastnerJSTAT10}, and this will also reflect in the final result for the microcanonical entropy derived in the next section.

\subsection{Asymptotic evaluation of the Laplace integral}
\label{s:Laplace}
In the canonical calculation by Tindemans and Capel \cite{TinCa74}, the multiple Laplace integral corresponding to \eqref{e:Omega3} is evaluated rigorously by constructing upper and lower bounds on the canonical free energy and showing that both bounds coincide with the maximum of the argument of the exponential function in the multiple Laplace integral. We did not succeed in adapting this strategy of proof to the microcanonical setting, mainly due to imaginary contributions in the exponent, which hamper the application of certain inequalities.

To avoid these difficulties, we evaluate the integral \eqref{e:Omega3} by a multi-dimensional version of the method of steepest descent (see \cite{Miller} for a textbook presentation), considering $N$ as the large parameter of the Laplace integral. This amounts to performing the thermodynamic limit $N\to\infty$ first, followed by the limit $n\to\infty$. Unfortunately we were not able to justify this exchange of the order of the two limiting procedures, so this step in the derivation (and also in Sec.\ 3.2.2 of \cite{KastnerJSTAT10}) is not rigorously justified.

To apply the method of steepest descent, we need to find a stationary point of the function $\mathcal{F}$ for which it is possible to smoothly deform the contours of the $s$- and $t$-integrations such that the paths of integration correspond to constant (zero) imaginary part of $\mathcal{F}$. Stationary points of $\mathcal{F}$ need to satisfy the conditions
\begin{subequations}
\begin{align}
0&=\frac{\partial\mathcal{F}}{\partial s} = e+\frac{1}{2ns^2}\sum_{i=1}^n\bm{x}^{(i)}\cdot\bm{x}^{(i)},\\
0&=\frac{\partial\mathcal{F}}{\partial t_\alpha} = m_\alpha+\frac{\tanh r}{r}\left(t_\alpha-\frac{\sqrt{\lambda_\alpha}}{n}\sum_{i=1}^n x_\alpha^{(i)}\right),\label{e:saddle1b}\\
0&=\frac{\partial\mathcal{F}}{\partial x_\alpha^{(u)}} = -\frac{x_\alpha^{(u)}}{ns}-\frac{\tanh r}{r} \frac{\sqrt{\lambda_\alpha}}{n}\left(t_\alpha-\frac{\sqrt{\lambda_\alpha}}{n}\sum_{i=1}^n x_\alpha^{(i)}\right),\label{e:saddle1c}
\end{align}
\end{subequations}
where $\alpha\in\{1,2,3\}$ and $u\in\{1,\dots,n\}$. Inserting \eqref{e:saddle1b} into \eqref{e:saddle1c} and some straightforward manipulations allow us to rewrite these equations as
\begin{subequations}
\begin{eqnarray}
0&=& 2nes^2+\sum_{i=1}^n\bm{x}^{(i)}\cdot\bm{x}^{(i)},\\
0&=& m_\alpha r+\left(t_\alpha-\frac{\sqrt{\lambda_\alpha}}{n}\sum_{i=1}^n x_\alpha^{(i)}\right)\tanh r,\label{e:saddle2b}\\
0&=& x_\alpha^{(u)}-m_\alpha s\sqrt{\lambda_\alpha}.\label{e:saddle2c}
\end{eqnarray}
\end{subequations}
In contrast to the derivation in \cite{KastnerJSTAT10} where a homogeneous solution, i.e.
\begin{equation}\label{e:xequal}
\bm{x}^{(u)} = \bm{x} = (x_1,x_2,x_3)\qquad\forall u\in\{1,\dots,n\},
\end{equation}
was postulated, we can now simply read off from \eqref{e:saddle2c} that $x_\alpha^{(u)}=m_\alpha s\sqrt{\lambda_\alpha}$ is independent of $u$, which implies that \eqref{e:xequal} must hold true for any stationary point of $\mathcal{F}$. Hence the stationary point equations simplify to
\begin{subequations}
\begin{eqnarray}
0&=& 2es^2+\bm{x}^2,\label{e:saddle3a}\\
0&=& m_\alpha r+\left(t_\alpha-\sqrt{\lambda_\alpha}x_\alpha\right)\tanh r,\label{e:saddle3b}\\
0&=& x_\alpha-m_\alpha s\sqrt{\lambda_\alpha},\label{e:saddle3c}
\end{eqnarray}
\end{subequations}
where
\begin{equation}\label{e:Rt1}
r=r(\bm{t},\bm{x}) = \sqrt{\sum_{\alpha=1}^3\left(t_\alpha-x_\alpha\sqrt{\lambda_\alpha}\right)^2}.
\end{equation}
For an asymptotic evaluation of the integrals in \eqref{e:Omega3} by means of the method of steepest descent, we have to evaluate $\mathcal{F}(s,\bm{t},\{\bm{x}^{(m)}\})$ as defined in \eqref{e:F} at the values of $s$, $\bm{t}$, and $\{\bm{x}^{(m)}\}$ specified by \eqref{e:saddle3a}--\eqref{e:saddle3c}. For $\mathcal{F}$ we obtain under condition \eqref{e:xequal} the expression
\begin{equation}\label{e:F2}
\mathcal{F}(s,\bm{t},\bm{x})=2es+\bm{m}\cdot\bm{t}-\frac{1}{2}\ln\left[1-\tanh^2 r(\bm{t},\bm{x})\right],
\end{equation}
where additionally \eqref{e:saddle3a} and the identity $2\ln\cosh x=-\ln(1-\tanh^2x)$ have been used. Making use of \eqref{e:saddle3a}--\eqref{e:saddle3c}, it is a matter of straightforward algebra to evaluate $\mathcal{F}$ at the values $s_0$, $\bm{t}_0$, and $\bm{x}_0$ which are solutions of these equations. The result is
\begin{equation}\label{e:Ffinal}
\mathcal{F}(s_0,\bm{t}_0,\bm{x}_0)=-|\bm{m}|\arctanh|\bm{m}|-\frac{1}{2}\ln\left(1-\bm{m}^2\right),
\end{equation}
and solutions of this type exist for all $\bm{m}$ satisfying $|\bm{m}|\leq1$ (see \ref{s:evalF} for a derivation).

According to the method of steepest descent, the asymptotic behavior of $\Omega_N$ in \eqref{e:Omega3} is now given as $\exp[N(\ln2+\mathcal{F})]$ times some prefactor (see for example Sec.\ 3.7 of Miller's textbook \cite{Miller} for the prefactor of multidimensional Laplace integrals, which can be adapted to the method of steepest descent of multidimensional integrals). The prefactor, however, is subexponential in $N$. Since we are interested in the microcanonical entropy \eqref{e:sem3def} in the thermodynamic limit, subexponential terms do not contribute and we obtain
\begin{equation}\label{e:sem}
s(\bm{m})=\ln 2 - |\bm{m}|\arctanh|\bm{m}|-\frac{1}{2}\ln\left(1-\bm{m}^2\right)
\end{equation}
as our final result for the microcanonical entropy of the anisotropic quantum Heisenberg model in the thermodynamic limit.

At first sight, this expression may appear independent of the energy $e$, but this is a matter of the viewpoint adopted. Inserting \eqref{e:saddle3c} into \eqref{e:saddle3a} gives the condition
\begin{equation}\label{e:em}
e=-\frac{1}{2}\sum_{\alpha=1}^3 \lambda_\alpha m_\alpha^2,
\end{equation}
indicating that the variables $e$ and $\bm{m}$ are overdetermining a macrostate of our model: Given all three components of the magnetization vector $\bm{m}$, the energy is already fixed. This implies that, in the four-dimensional parameter space $(e,\bm{m})$, the entropy lives only on a three-dimensional submanifold. From a geometric point of view, equations \eqref{e:sem} and \eqref{e:em} can be interpreted as follows: In the three-dimensional parameter space $(m_1,m_2,m_3)$, the entropy \eqref{e:sem} is a central symmetric function. For a fixed value of the energy $e$, condition \eqref{e:em} defines an ellipsoid in this space.

Because of its symmetry properties, the entropy \eqref{e:sem} has a particularly simple appearance. However, for what will be discussed in later sections of this article, it is convenient to rewrite $s$ as a function of energy $e$ and two of the magnetization components. To this purpose we solve \eqref{e:em} for $m_1$, yielding
\begin{equation}
m_1(e,m_2,m_3)=\sqrt{\frac{-2e - \lambda_2 m_2^2 - \lambda_3 m_3^2}{\lambda_1}}
\end{equation}
and
\begin{equation}\label{e:m_of_emm}
|\bm{m}(e,m_2,m_3)|=\sqrt{\frac{1}{\lambda_1}[(\lambda_1 - \lambda_2)m_2^2 + (\lambda_1 - \lambda_3)m_3^2 - 2e]},
\end{equation}
where we have assumed $\lambda_1\neq0$.\footnote{If $\lambda_1$ happens to be zero, one chooses instead to solve for a magnetization component $m_\alpha$ corresponding to a nonzero $\lambda_\alpha$.} Inserting the latter expression into \eqref{e:sem}, the entropy $s(e,m_2,m_3)$ is obtained.\footnote{With a slight abuse of notation, we use here the same symbol $s$ for different entropy functions.} In the prequel \cite{KastnerJSTAT10} to this article, an entropy $s(e,m_3)$ was derived as a function of only two variables, namely the energy and the $3$-component of the magnetization. This result can be recovered from $s(e,m_2,m_3)$ by contraction with respect to $m_2$,
\begin{equation}\label{e:s_red}
s(e,m_3)=\max_{m_2} s(e,m_2,m_3).
\end{equation}

\subsection{Properties of the microcanonical entropy}
\label{s:properties}
Among the many properties of the microcanonical entropy, we want to focus in particular on whether or not it is a concave function. This property, as alluded to in the Introduction, is crucial for determining whether or not statistical ensembles are equivalent: Finding a nonconcave microcanonical entropy will tell us that the corresponding canonical ensemble are nonequivalent. For a function $f$ of several variables, concavity means that
\begin{enumerate}
\item The domain $\mathcal{D}$ of $f$ is a convex set, i.e.\ $(1-\lambda)\bm{x} + \lambda \bm{x}' \in\mathcal{D}$ whenever $\bm{x},\bm{x}'\in\mathcal{D}$ and $\lambda\in[0,1]$.
\item For all $\bm{x},\bm{x}'\in\mathcal{D}$ and $\lambda\in(0,1)$, we have
\begin{equation}\label{e:concave}
f((1-\lambda)\bm{x} + \lambda \bm{x}') \geq (1-\lambda) f(\bm{x}) + \lambda f(\bm{x}').
\end{equation}
\end{enumerate}

The concavity properties will crucially depend on whether we consider the microcanonical entropy \eqref{e:sem} as a function of $(m_1,m_2,m_3)$, $(e,m_2,m_3)$, or the reduced entropy \eqref{e:s_red} as a function of $(e,m_3)$. The latter has been studied in detail in \cite{KastnerJSTAT10}, finding that $s(e,m_3)$ is a concave function if at least one of the coupling constants $\lambda_1$, $\lambda_2$ is larger than $\lambda_3$. If, however, $\lambda_3$ is the largest coupling constant, then the microcanonical entropy $s(e,m_3)$ is nonconcave, indicating nonequivalence with the canonical ensemble with the inverse temperature $\beta$ and the 3-component $h_3$ of the magnetic field as control parameters. 

For the more general entropy \eqref{e:sem} derived in the present article, the concavity properties are most easily discussed when considering $s$ as a function of the magnetization components $(m_1,m_2,m_3)$. As is evident from \eqref{e:sem}, the entropy depends only on the modulus $|\bm{m}|$ of the magnetization vector. Calculating the second derivative
\begin{equation}
\frac{\partial^2 s(|\bm{m}|)}{\partial |\bm{m}|^2} = \frac{1}{\bm{m}^2-1},
\end{equation}
we find that this second derivative is negative on the entire domain $|\bm{m}|\in[0,1]$. By symmetry, this implies that $s(m_1,m_2,m_3)$ is a concave function on its domain
\begin{equation}
\mathcal{D}_{\bm{m}}=\left\{\bm{m}\in\RR^3\,\big|\,|\bm{m}|\leq1\right\}.
\end{equation}

\begin{figure}\center
\includegraphics[width=0.3\linewidth]{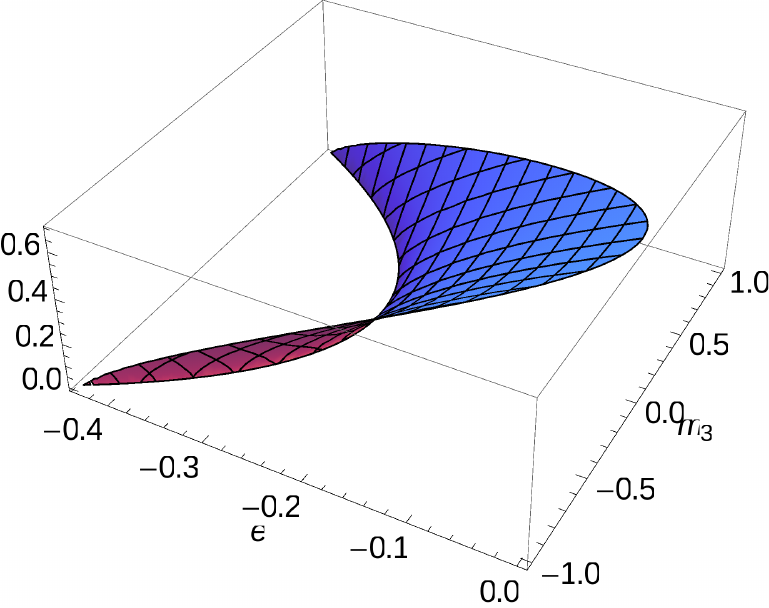}
\hspace{0.08\linewidth}
\includegraphics[width=0.3\linewidth]{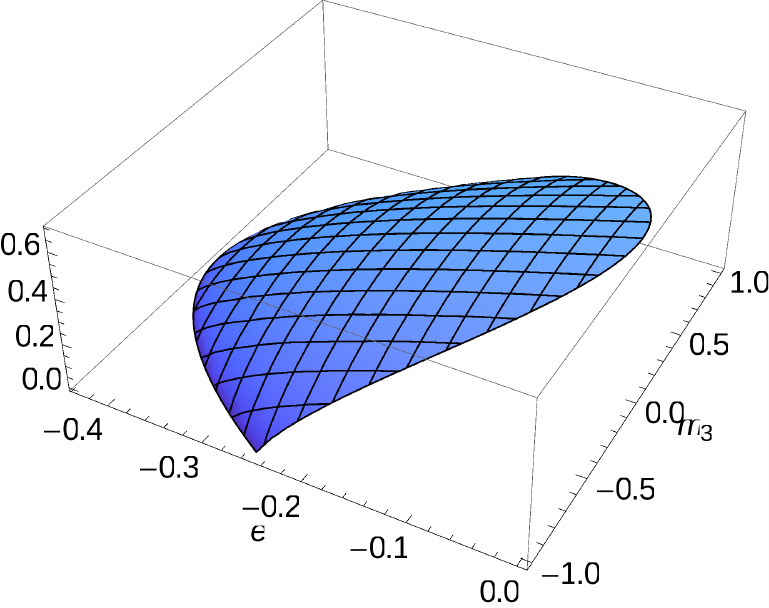}
\caption{\label{f:sem_const}
Graphs of the entropy $s(e,0,m_3)$, i.e.\ for one variable fixed to zero. Whether the resulting two-variable function is concave or not depends on whether $\lambda_1/\lambda_3$ is smaller or greater than 1. This behavior is illustrated for $(\lambda_1,\lambda_2,\lambda_3)=(1/2,0,\sqrt{3}/2)$ (left) and $(\lambda_1,\lambda_2,\lambda_3)=(\sqrt{3}/2,0,1/2)$ (right).
}
\end{figure}

The situation changes when the microcanonical entropy is considered as a function of the energy $e$ and two of the magnetization components, say $m_2$ and $m_3$. Graphically, we are limited to plotting the entropy as a function of two variables, but we will see that this is sufficient to get an idea of the concavity properties of $s(e,m_2,m_3)$. Equation \eqref{e:concave} defines concavity by comparing the function values at two points in $\mathcal{D}$ to the function values at all points on a straight line connecting these two endpoints. To show that $s(e,m_2,m_3)$ can be nonconcave, it is therefore sufficient to fix one of the variables (say, $m_2$), and investigate $s$ in dependence of the remaining two variables. A nonconcavity in the two-variable function $s(e,m_2,m_3)|_{m_2=\mbox{\footnotesize const.}}$ then implies that also $s(e,m_2,m_3)$ is nonconcave. The plots in Fig.\ \ref{f:sem_const} illustrate that, for certain choices of the coupling constants, $s(e,m_2,m_3)$ is indeed a nonconcave function. Interestingly, although the entropies $s(\bm{m})$ and $s(e,m_2,m_3)$ describe the same physical situation of fixed energy $e$ and magnetization vector $\bm m$, the former is a concave function, while the latter is not.

We conclude this survey of properties of the microcanonical entropy by discussing the microcanonical magnetic susceptibility in the $i$-direction at constant energy,
\begin{equation}\label{e:chi}
\chi_i (e,m_2,m_3)  = \left(\frac{\partial h}{\partial m_i}\right)^{-1}
= \left(\frac{\partial s}{\partial e} \right)^2 \left( \frac{\partial^2 s}{\partial m_i \partial e} \frac{\partial s}{\partial m_i} - \frac{\partial s}{\partial e} \frac{\partial^2 s}{\partial m_i^2} \right)^{-1},
\end{equation}
where $i\in\{2,3\}$. This formula is a straightforward modification of equation (61) in \cite{CaRuTou07}, obtained by replacing $m$ by one of the magnetization components $m_i$. A negative magnetic susceptibility, though usually considered a hallmark of nonequivalent ensembles, is a sufficient, but not a necessary condition for the onset of a nonconcavity of $s$ (see Sec.\ 5.3 of \cite{CaRuTou07} for a discussion). Evaluating \eqref{e:chi} by making use of \eqref{e:sem} and \eqref{e:m_of_emm}, we obtain
\begin{equation}
\chi_i = \frac{1}{\lambda_1-\lambda_i}.
\end{equation}
Remarkably, $\chi_i$ is independent of $e$, $m_2$, and $m_3$, so the magnetic susceptibility does not even depend on the macrostate the system is in. Moreover, we observe that the sign of $\chi_i$ depends only on whether $\lambda_1/\lambda_i$ is smaller or greater than 1. Hence, for the model studied here, the magnetic susceptibility becomes negative precisely when the microcanonical entropy develops a nonconcavity.

\subsection{Comparison with earlier results}
The results of Sec.\ \ref{s:sem}, although for a more general model, are to a certain extent similar to the ones reported in \cite{KastnerJSTAT10}. Before moving on to a different, and more original, topic in Sec.\ \ref{s:shu}, we briefly want to summarize and highlight the differences of the results.
\begin{enumerate}
\item The external magnetic field $\bm h$ in the Hamiltonian \eqref{e:Hamiltonian} is not restricted to the $3$-direction, but can be orientated in any spatial direction.
\item The microcanonical entropy $s(e,\bm m)$ is calculated as a function of the energy $e$ and the magnetization vector $\bm m=(m_1,m_2,m_3)$, as compared to $s(e,m_3)$ in \cite{KastnerJSTAT10}.
\item The microcanonical density of states \eqref{e:OmegaDef} is defined to be the inverse Laplace transform of the canonical partition function, which is conceptually clearer than the definition advocated in \cite{KastnerJSTAT10}, but leads to identical results in the thermodynamic limit.
\item Whereas in \cite{KastnerJSTAT10} a homogeneous solution \eqref{e:xequal} of the Laplace integral \eqref{e:Omega3} was postulated, this is now proved to be the unique solution.
\item Expressed in terms of the magnetization vector $\bm m$, the microcanonical entropy is a more symmetric, and therefore simpler, function than the one reported in \cite{KastnerJSTAT10}.
\end{enumerate}
The results of Ref.\ \cite{KastnerJSTAT10} and the present paper both share the shortcoming of two nonrigorous steps in the derivation of the microcanonical entropy, as commented on in Secs.\ \ref{s:trace} and \ref{s:Laplace}. A related calculation of the microcanonical entropy of a long-range spin model, yielding essentially the same result, has been reported in \cite{Mori12}. However, according to our understanding of that paper, the microcanonical proof is also incomplete, as the validity of certain mappings is demonstrated only in the canonical context. This leaves the rigorous proof of \eqref{e:sem} as an open problem.

\section{Ensemble of fixed energy and magnetic field}
\label{s:shu}
The microcanonical entropies $s(\bm m)$ and $s(e,m_2,m_3)$ discussed in this article describe the physical situation of fixed energy $e$ and fixed magnetization $\bm m$. It is not immediately obvious how these constraints can be realized in experiment: The quantum Heisenberg model was devised to model ferromagnetic spin systems which, in their traditional condensed matter realizations, are typically coupled to a thermal reservoir. As a consequence, the energy is not fixed, but fluctuates around a certain mean value, and the canonical ensemble is appropriate for a statistical equilibrium description of this situation.

Alternatively, quantum Heisenberg models can be emulated experimentally by means of ultracold atoms or molecules in optical lattices, or by trapped ion crystals. All these realizations of condensed matter-type systems by ultracold gases possess the attractive feature of being highly controllable: the interaction type and strength can be tuned, and even the character of the interaction force can be switched from attractive to repulsive. The total energy and number of particles in these experiments are conserved to a very good degree and, as a consequence, a statistical description of the equilibrium properties of such systems should make use of the microcanonical ensemble. The total magnetization in such systems, however, is usually not a conserved quantity. As a consequence, the statistical ensemble realized in such an experimental setting is a microcanonical one with constant energy, but fluctuating magnetization (also called a {\em mixed ensemble} in \cite{ElHaTur00}).

To account for this experimental situation, we switch to a statistical ensemble in which the total energy\footnote{The reader be reminded that, according to \eqref{e:OmegaDef}, the energy $e$ was defined as the interaction energy per spin, but did not include the energy contribution originating from the Zeeman term in the Hamiltonian \eqref{e:Hamiltonian}. The total energy (per spin) $u$ is defined such that it accounts for both contributions.}
\begin{equation}\label{e:u}
u=e-\bm{h}\cdot\bm{m}=-\frac{1}{2}\sum_{\alpha=1}^3 \lambda_\alpha m_\alpha^2-\sum_{\alpha=1}^3 h_\alpha m_\alpha=-\sum_{\alpha=1}^3 m_\alpha\left(\frac{\lambda_\alpha m_\alpha}{2}+ h_\alpha\right)
\end{equation}
is fixed, while the magnetization is allowed to fluctuate. In the following, we will use the information contained in the generalized microcanonical entropy $s(e,\bm m)$ to compute, for a given fixed magnetic field $\bm h$, the microcanonical entropy $s_{\bm{h}}(u)$ at constant total energy $u$.

\subsection{Derivation of \texorpdfstring{$s_{\bm{h}}(u)$}{sh(u)}}
\label{s:shu_derivation}
The strategy for the computation of $s_{\bm{h}}(u)$ is as follows: We solve \eqref{e:u} for one of the magnetization components, say
\begin{equation}
m_1(u,m_2,m_3)=-\frac{1}{\lambda_1}\left(h_1\pm\sqrt{h_1^2 - \lambda_1\left(2u+\lambda_2m_2^2+\lambda_3m_3^2+2h_2m_2+2h_3m_3\right)}\right).
\end{equation}
Inserting this expression into the generalized microcanonical entropy $s(\bm m)$ in \eqref{e:sem}, we obtain an entropy $\tilde{s}_{\bm h}(u,m_2,m_3)$ that depends on the total energy $u$, two magnetization components $m_2$ and $m_3$, and on all three components of the magnetic field $\bm h$. Since the magnetization is allowed to fluctuate, entropy will drive the system towards those values of $m_2$ and $m_3$ for which $\tilde{s}$ is maximized. Hence, the microcanonical entropy
\begin{equation}\label{e:smax}
s_{\bm{h}}(u)=\max_{m_2,m_3}\tilde{s}_{\bm h}(u,m_2,m_3)
\end{equation}
describes, for a given magnetic field $\bm h$, the system at fixed total energy $u$. The maximization in \eqref{e:smax} can be achieved by finding the stationary solutions $m_2(u)$ and $m_3(u)$ which, simultaneously for $j=2,3$, satisfy the equations
\begin{equation}
0=\frac{\partial \tilde{s}_{\bm h}(u,m_2,m_3)}{\partial m_j}=-\frac{\arctanh|\bm m(u,m_2,m_3)|}{2|\bm m(u,m_2,m_3)|}\frac{\partial \bm m^2(u,m_2,m_3)}{\partial m_j},
\end{equation}
where
\begin{equation}
\bm m(u,m_2,m_3)=\begin{pmatrix}m_1(u,m_2,m_3)\\m_2\\m_3\end{pmatrix}.
\end{equation}
Since $\arctanh(x)/x\geq1$, this amounts to solving
\begin{equation}\label{e:dm2dmj}
0=\frac{\partial \bm m^2(u,m_2,m_3)}{\partial m_j},\qquad j=2,3,
\end{equation}
for $m_2$ and $m_3$. Inserting these solutions $m_2(u,\bm h)$ and $m_3(u,\bm h)$ into $\tilde{s}_{\bm h}(u,m_2,m_3)$, the microcanonical entropy $s_{\bm{h}}(u)$ is obtained.

\subsection{\texorpdfstring{$s_{\bm{h}}(u)$}{sh(u)} for special parameter values}
\label{s:shu_properties}
\begin{figure}\center
\includegraphics[height=0.26\linewidth]{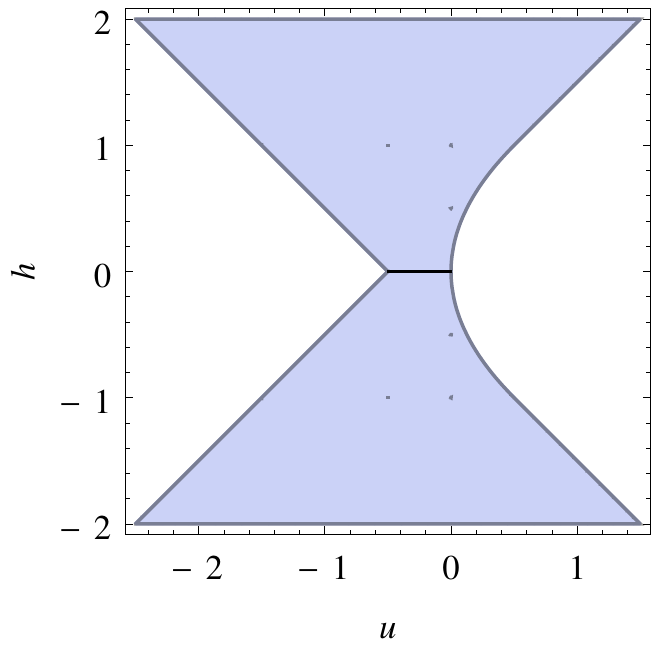}
\hfill
\includegraphics[height=0.25\linewidth]{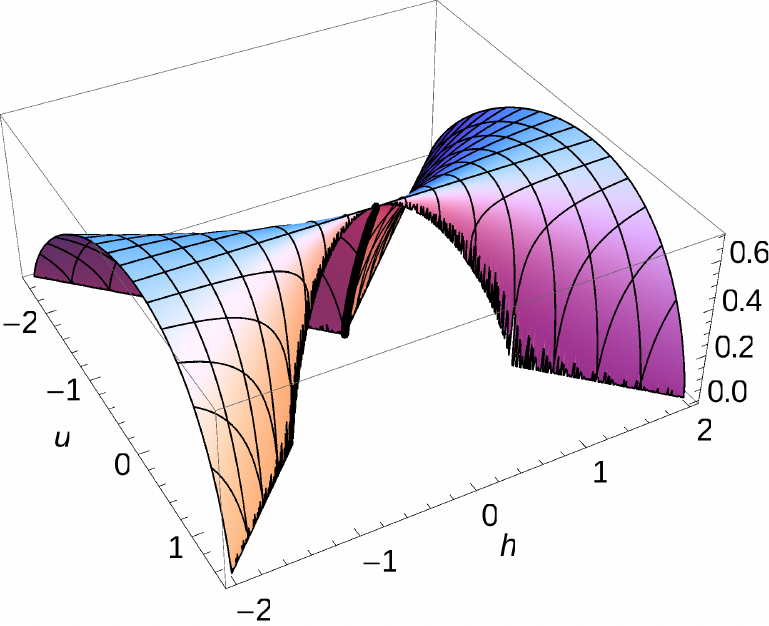}
\hfill
\includegraphics[height=0.25\linewidth]{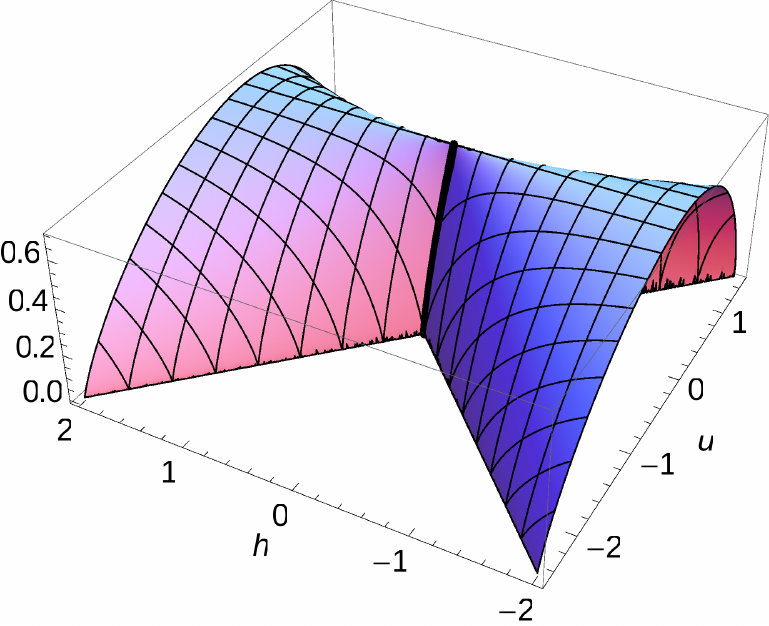}
\newline
\includegraphics[height=0.26\linewidth]{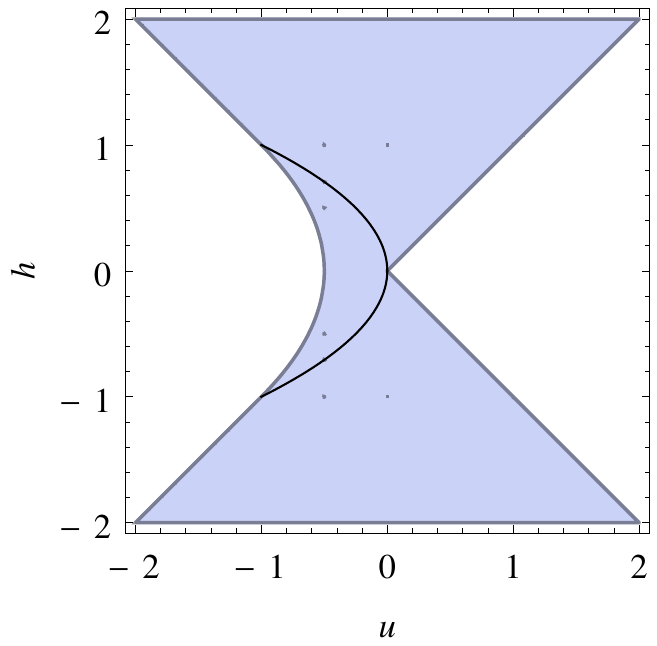}
\hfill
\includegraphics[height=0.25\linewidth]{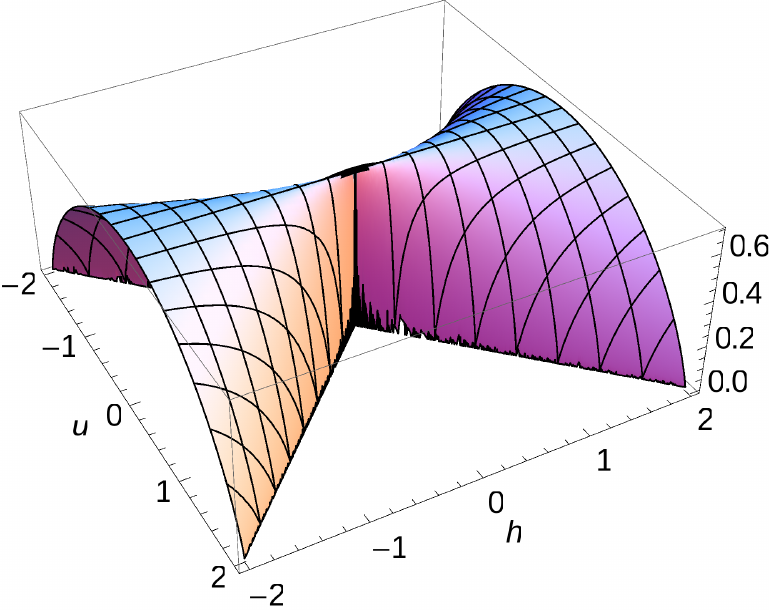}
\hfill
\includegraphics[height=0.25\linewidth]{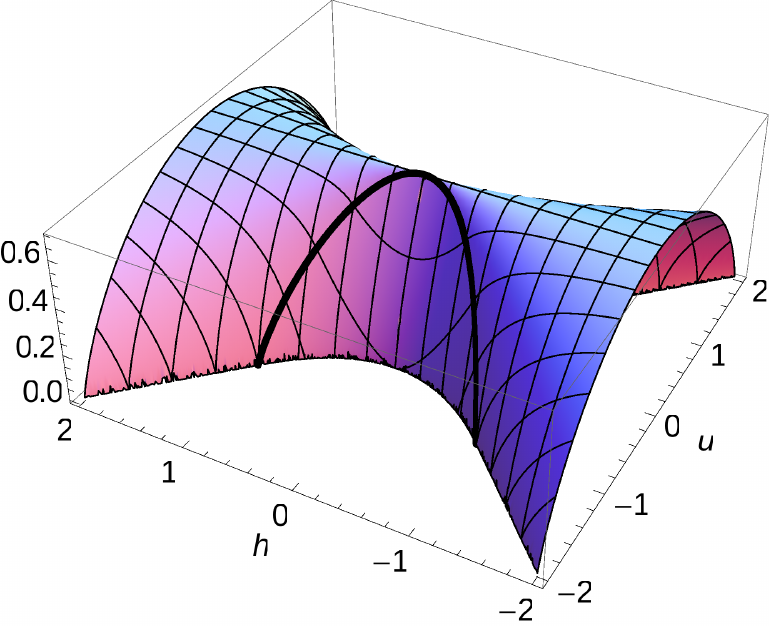}
\caption{\label{f:iso_isi}
For a magnetic field $\bm h=(h,0,0)$ in $x$-direction, the graph of the microcanonical entropy $s_{\bm h}(u)$ is shown as a function of the total energy $u$ and the magnitude of the magnetic field strength $h$. The three plots in the top row are for the Ising model in a longitudinal field ($\lambda_1=1$, $\lambda_2=\lambda_3=0$), or for the isotropic Heisenberg model ($\lambda_1=\lambda_2=\lambda_3=1$) whose entropies are identical. The plots in the bottom row are for the Ising model in a transverse field ($\lambda_1=\lambda_3=0$ and $\lambda_2=1$). The hatched areas in the two-dimensional plots show the regions in the $(u,h)$-plane for which the entropy is defined, the black lines in the interior of these regions indicate nonanalyticities of the entropy.
}
\end{figure}

It is instructive to first discuss the entropy $s_{\bm{h}}(u)$ for three particularly simple special cases: the Ising model in a longitudinal magnetic field, the Ising model in a transverse magnetic field, and the isotropic Heisenberg model. We refrain from presenting the analytic formul{\ae} resulting from the evaluation of \eqref{e:dm2dmj}. Instead we show in Fig.\ \ref{f:iso_isi} exemplary plots of the entropy functions in the $(u,h)$-plane, where $h$ is the magnitude of a magnetic field $\bm h=(h,0,0)$ in $x$-direction. We will discuss these plots in some detail, as variations and deformations of the features they display will show up also in the case of more general parameter values discussed in Sec.\ \ref{s:shu_general}. 

For the Ising model in a longitudinal field ($\lambda_1=1$, $\lambda_2=\lambda_3=0$), the entropy is shown in the top row of Fig.\ \ref{f:iso_isi}. Nonanalytic behavior occurs along the $h=0$ line, and this corresponds to a field-driven transition from a phase of positive magnetization to one of negative magnetization. Considering the entropy $s_{\bm h}(u)$ along a slice of constant $h$ in the second plot of Fig.\ \ref{f:iso_isi}, it is a concave function of $u$ for any fixed value of $h$. Because of concavity in $u$, and since $u$ is the only microcanonical variable, we conclude that the microcanonical ensemble of constant total energy $u$ is equivalent to the canonical ensemble in which temperature is the variable conjugate to $u$. Along the line $h=0$, the entropy shows a peculiar behavior, as it is a strictly monotonous function of $u$, terminating at $(u,h)=(0,0)$ with positive slope. Such a behavior was discussed in detail in \cite{CaKa07} under the name of {\em partial equivalence of ensembles}\/ (see in particular Figs.\ 4 and 6 of \cite{CaKa07}). In this case, the endpoint at $(0,0)$ corresponds to the zero-field phase transition from a ferromagnetic to a paramagnetic phase. Remarkably, the single microcanonical macrostate at $(0,0)$ coincides with the canonical macrostates for {\em all} temperatures larger than the Curie temperature. For any nonzero $h$, however, the entropy bends down to negative slopes, similar to what was reported in Fig.\ A1 of \cite{CaKa07}.

For the isotropic Heisenberg model, the entropy is identical to that of the Ising model in a longitudinal field (top row of Fig.\ \ref{f:iso_isi}).

For the Ising model in a transverse magnetic field ($\lambda_1=\lambda_3=0$ and $\lambda_2=1$), the entropy is shown in the bottom row of Fig.\ \ref{f:iso_isi}. Nonanalytic behavior occurs along the parabola $u=-h^2$ in the $(u,h)$-plane. Inside the crescent-shaped region bounded by this parabola, the system is in a ferromagnetic phase, characterized by a nonvanishing $3$-component of the magnetization. In the hatched region outside this crescent-shape, the magnetization is fully orientated in the field-direction ($1$-direction). As for the case of a longitudinal field, the entropy is again concave in $u$ for any fixed value of $h$.

\subsection{General properties of \texorpdfstring{$s_{\bm{h}}(u)$}{sh(u)}}
\label{s:shu_general}
For more general parameter values, the entropy $s_{\bm{h}}(u)$ shows a combination of the features we observed for the special cases in Fig.\ \ref{f:iso_isi}, i.e.\ both, the field-driven straight transition line at $h=0$ and the arc-shaped transition line of the transverse-field Ising model are present. The shapes and sizes of the different regions (or phases) change upon variation of the parameters. There are in principle six parameters to manipulate ($\lambda_1$, $\lambda_2$, $\lambda_3$, $h_1$, $h_2$, and $h_3$), although one of them can be fixed without loss of generality (since an overall prefactor in the Hamiltonian only leads to a trivial rescaling of the units of energy). Here we will be analyzing two different submanifolds in parameter space.

\begin{figure}\center
\includegraphics[height=0.2\linewidth]{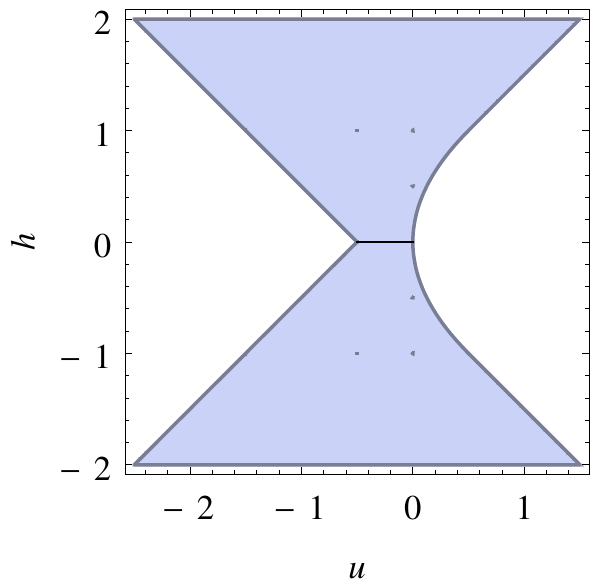}
\includegraphics[height=0.19\linewidth]{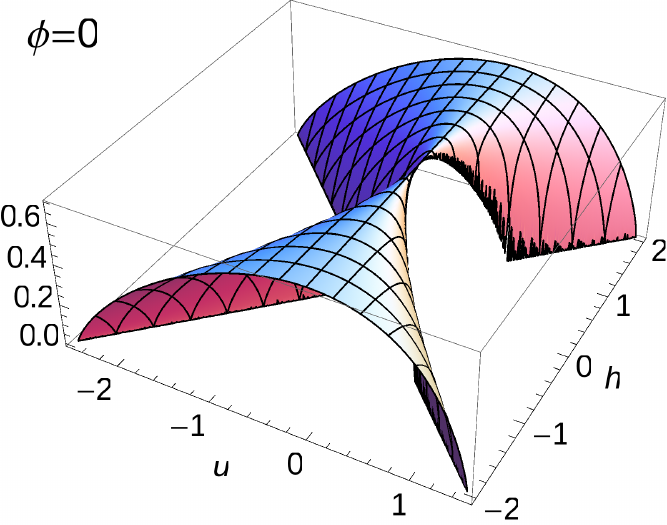}
\hfill
\includegraphics[height=0.2\linewidth]{shuPhi0domain.pdf}
\includegraphics[height=0.19\linewidth]{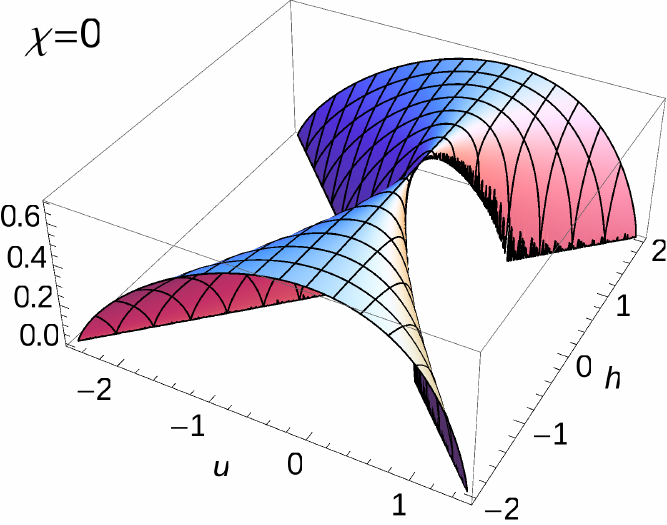}
\newline
\includegraphics[height=0.2\linewidth]{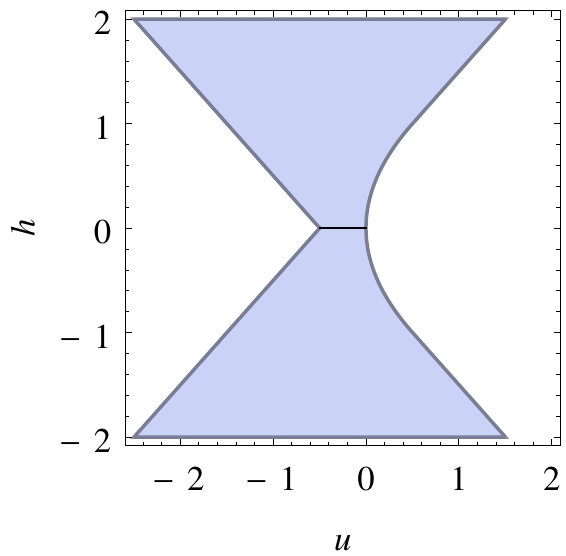}
\includegraphics[height=0.19\linewidth]{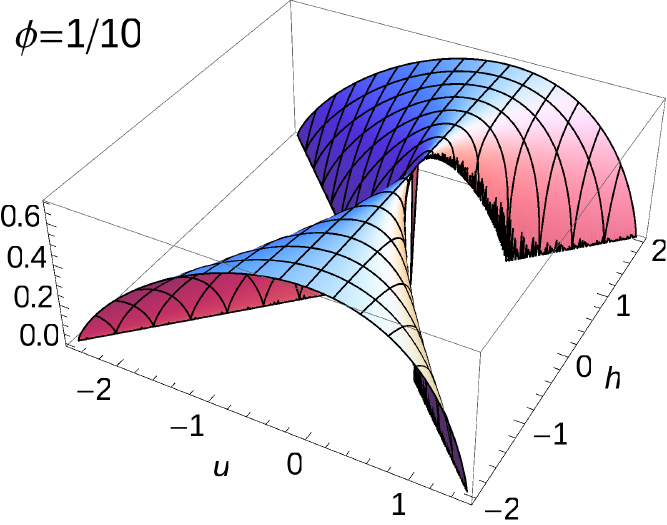}
\hfill
\includegraphics[height=0.2\linewidth]{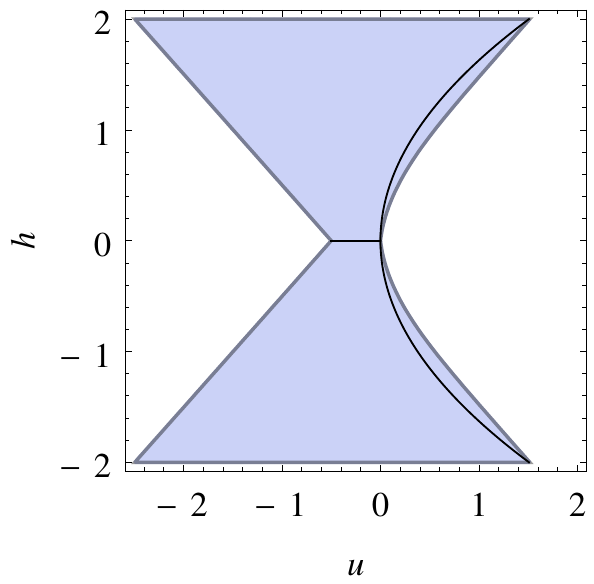}
\includegraphics[height=0.19\linewidth]{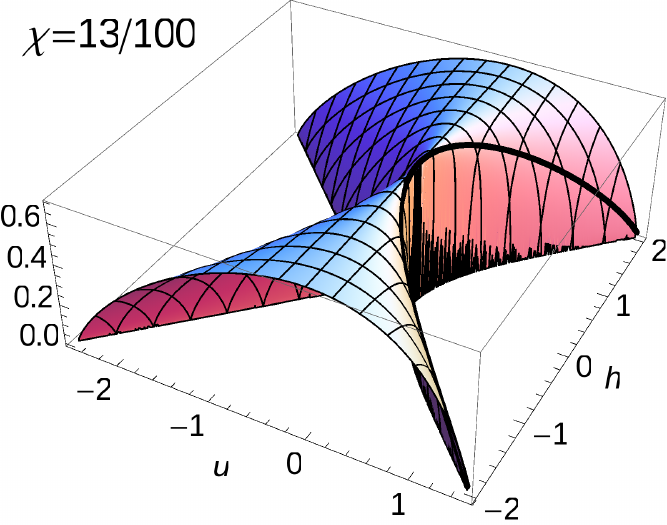}
\newline
\includegraphics[height=0.2\linewidth]{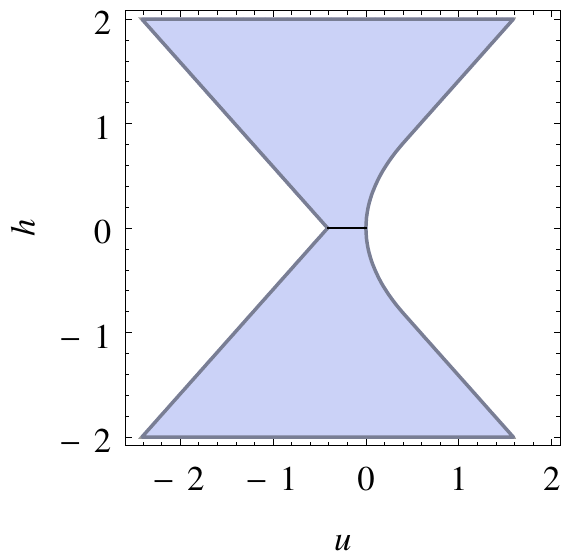}
\includegraphics[height=0.19\linewidth]{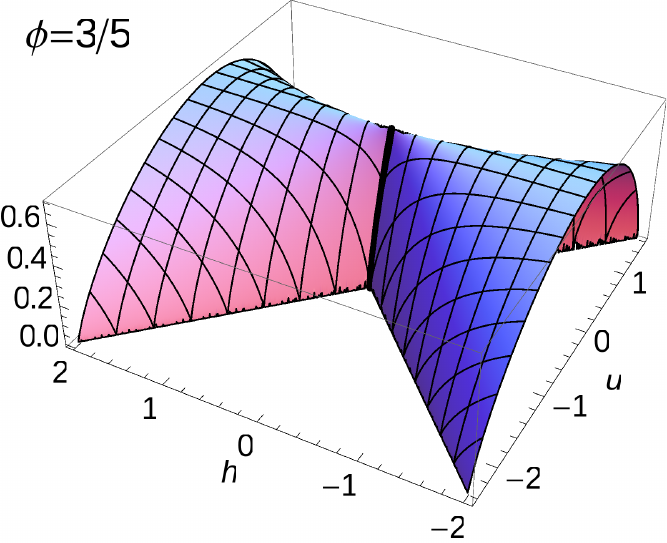}
\hfill
\includegraphics[height=0.2\linewidth]{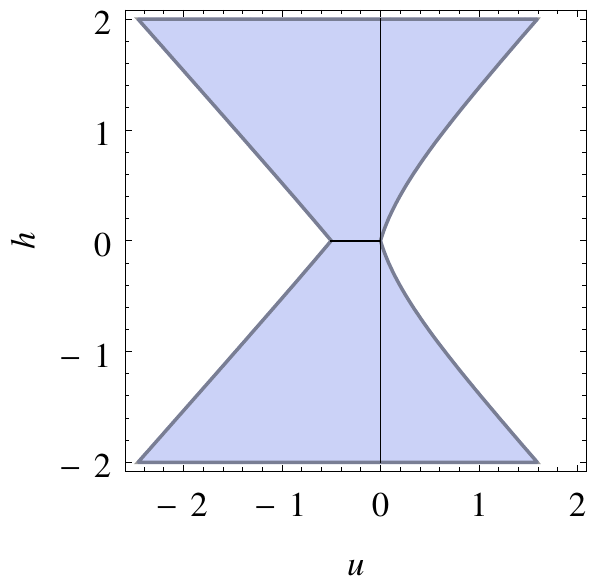}
\includegraphics[height=0.19\linewidth]{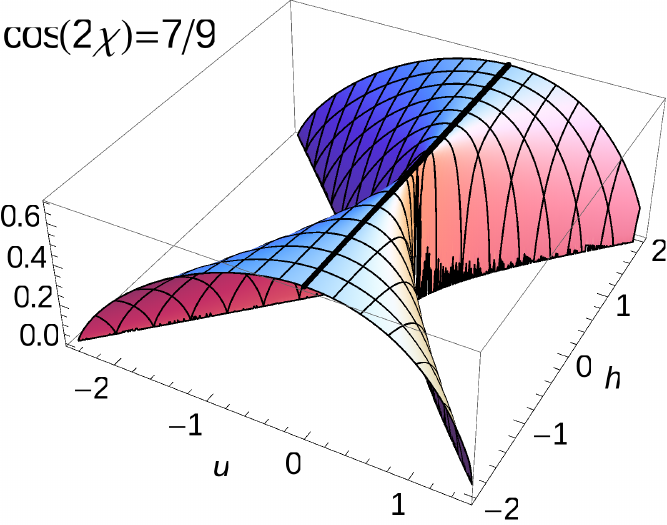}
\newline
\includegraphics[height=0.2\linewidth]{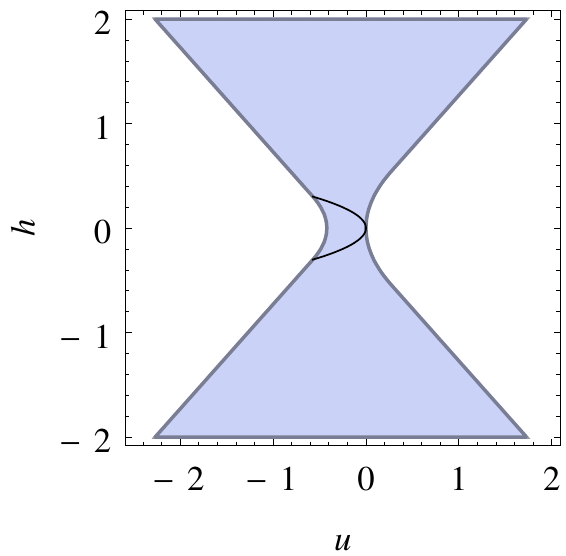}
\includegraphics[height=0.19\linewidth]{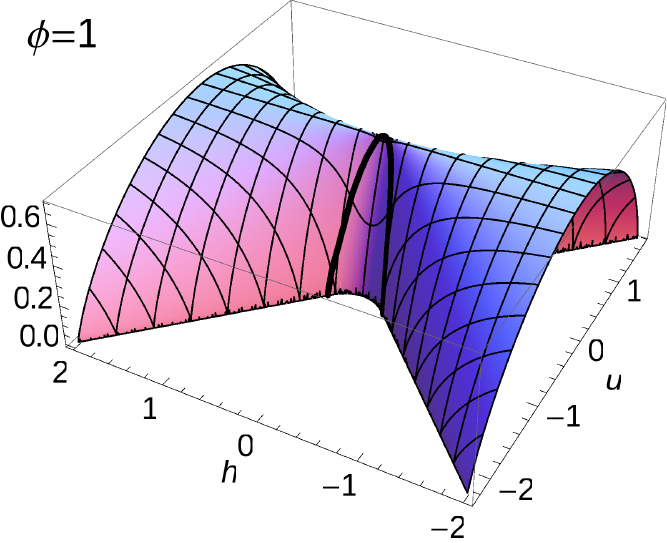}
\hfill
\includegraphics[height=0.2\linewidth]{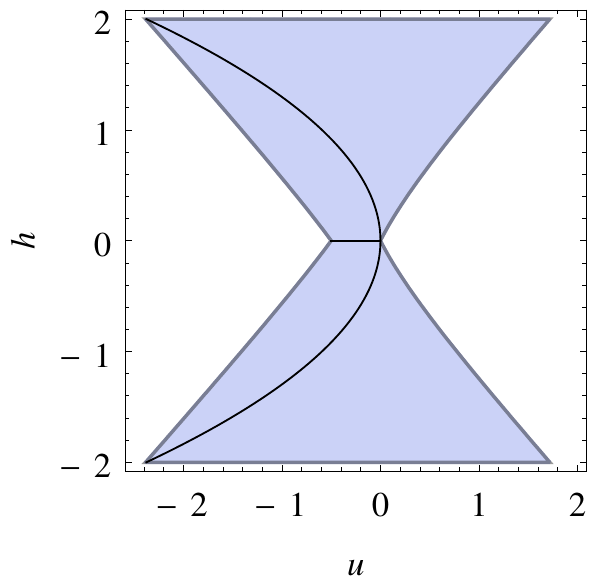}
\includegraphics[height=0.19\linewidth]{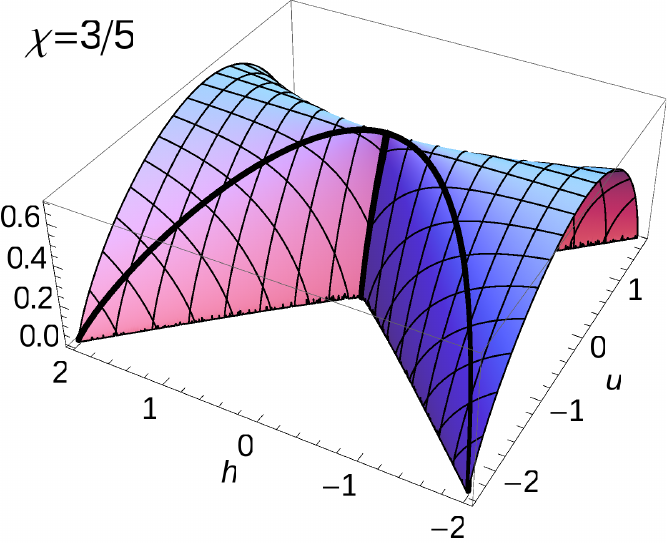}
\newline
\includegraphics[height=0.2\linewidth]{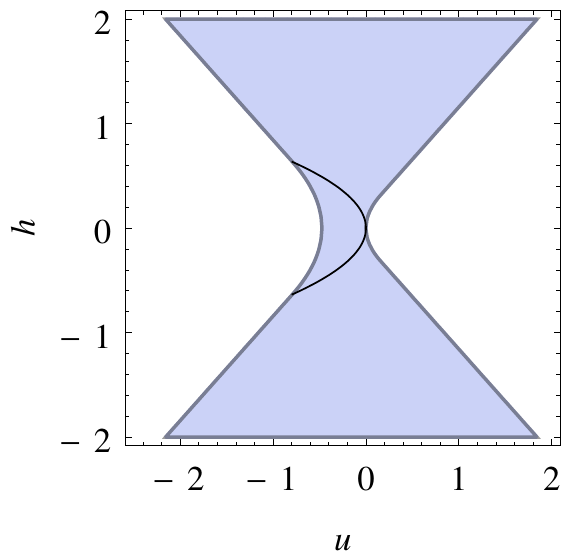}
\includegraphics[height=0.19\linewidth]{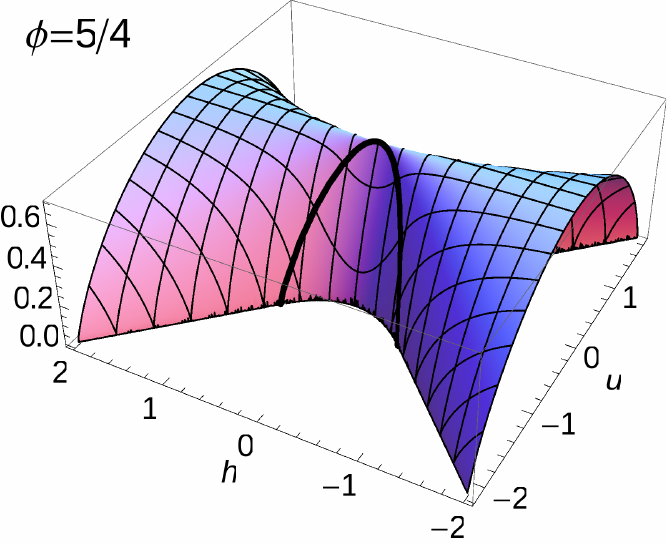}
\hfill
\includegraphics[height=0.2\linewidth]{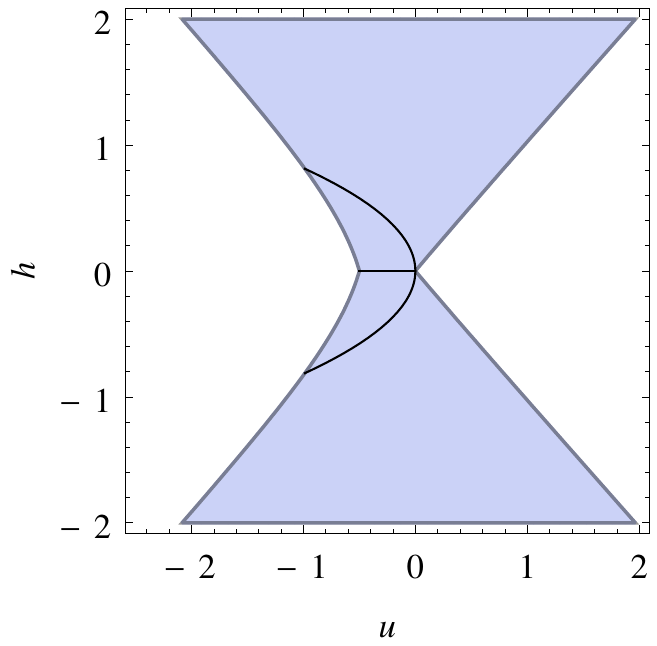}
\includegraphics[height=0.19\linewidth]{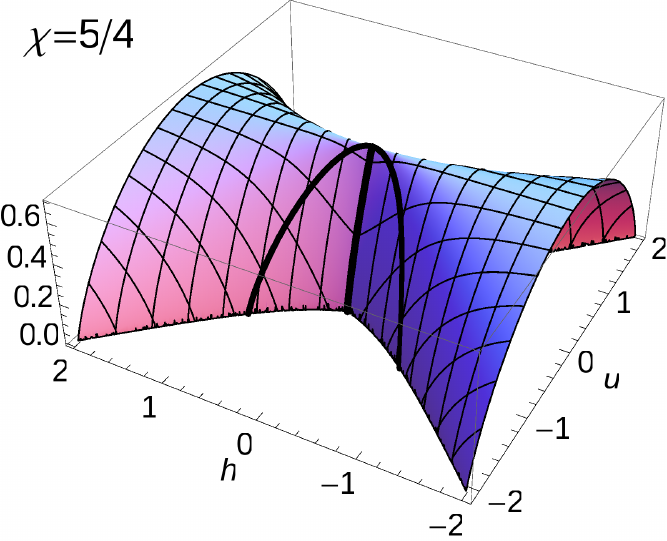}
\newline
\includegraphics[height=0.2\linewidth]{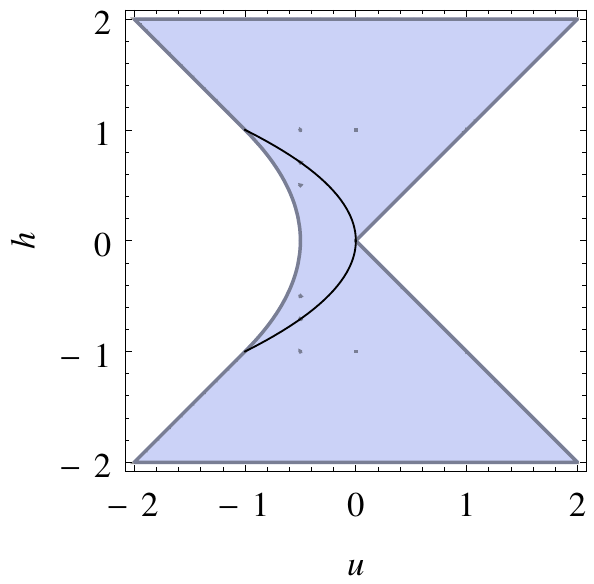}
\includegraphics[height=0.19\linewidth]{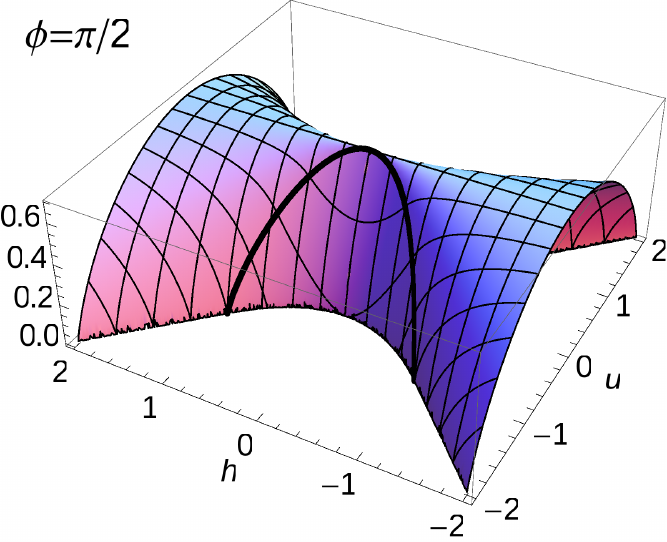}
\hfill
\includegraphics[height=0.2\linewidth]{shuPhiPi2domain.pdf}
\includegraphics[height=0.19\linewidth]{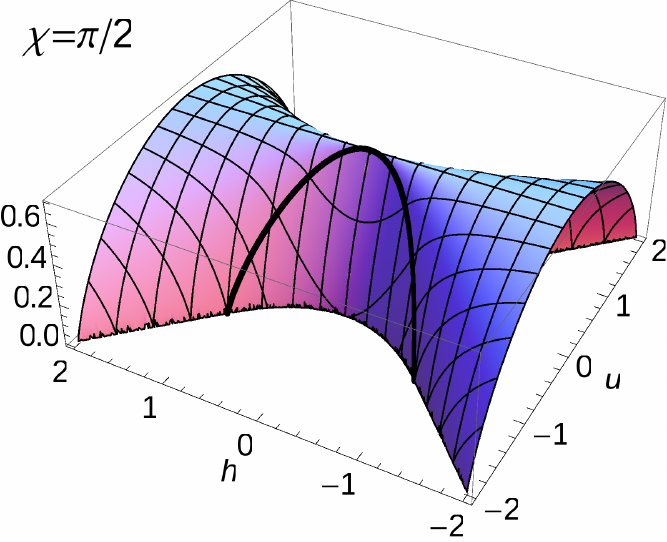}
\caption{\label{f:shu_general}
Plots of the microcanonical entropy $s_{\bm h}(u)$ for the field orientations and coupling strengths described in Sec.\ \ref{s:shu_general}. Left two columns: for a magnetic field oriented along the 1-direction and spin--spin couplings $\lambda_1=\cos\phi$, $\lambda_2=\sin\phi$ in the (1,2)-plane, with angles $\phi=0$, $1/10$, $3/5$, $1$, $5/4$, and $\pi/2$ (from top to bottom). The hatched areas in the two-dimensional plots show the regions in the $(u,h)$-plane for which the entropy is defined, the black lines in the interior of these regions indicate nonanalyticities of the entropy. Right two columns: as on the left, but for spin--spin couplings along the 1-direction and a magnetic field oriented in the (1,2)-plane, with angles $\chi=0$, $13/100$, $\arccos(7/9)/2$, $3/5$, $5/4$, and $\pi/2$ (from top to bottom).
}
\end{figure}

\subsubsection{Coupling in the \texorpdfstring{$(1,2)$}{(1,2)}-plane, field in the \texorpdfstring{$1$}{1}-direction.}
\label{s:1scenario}
The first submanifold in $(\boldsymbol\lambda,\bm h)$ space is parametrized by an angle $\phi$ and by the modulus $h=|\bm h|$ of the magnetic field, fixing the parameter values to $\lambda_1=\cos\phi$, $\lambda_2=\sin\phi$, $\lambda_3=0$, $h_1=h$, and $h_2=h_3=0$. This choice corresponds to a magnetic field oriented along the 1-direction and spin--spin couplings in the (1,2)-plane. 
Plots of the entropy $s_{\bm{h}}(u)$ are shown in Fig.\ \ref{f:shu_general} (left two columns) for various values of the angle $\phi$. The change of behavior is best understood by considering the plots from bottom to top, i.e.\ by starting from the transverse-field Ising model ($\phi=\pi/2$) and monitoring the changes upon rotation of the field into the longitudinal direction ($\phi=0$). In this sequence of plots, the bow-shaped transition line of the transverse-field Ising model is narrowing to a hairpin, and eventually, around $\phi=\pi/4$, collapsing into the zero-field transition line of the Ising model in a longitudinal field.

\subsubsection{Coupling in the \texorpdfstring{$1$}{1}-direction, field in the \texorpdfstring{$(1,2)$}{(1,2)}-plane.}
\label{s:2scenario}
The second submanifold in $(\boldsymbol\lambda,\bm h)$ parameter space we consider is parametrized by an angle $\chi$ and by the modulus $h=|\bm h|$ of the magnetic field, fixing the parameter values to $\lambda_1=1$, $\lambda_2=\lambda_3=0$, $h_1=h\cos\chi$, $h_2=h\sin\chi$, and $h_3=0$. This choice corresponds to a spin--spin coupling along the 1-direction and a magnetic field oriented in the (1,2)-plane. Plots of the entropy $s_{\bm{h}}(u)$ are shown in Fig.\ \ref{f:shu_general} (right two columns) for various values of the angle $\chi$. The special cases $\chi=0$ (top row; coupling is aligned with the magnetic field) and $\chi=\pi/2$ (bottom row; coupling transverse to the magnetic field) are identical to those in the preceding paragraph, but this is not the case for intermediate angles. Starting again from the transverse-field Ising model in the bottom row, we observe that, as soon as the angle deviates from the transverse-field value of $\chi=\pi/2$, the zero-field transition line (familiar from the Ising model in a longitudinal field) pops up. The bow-shaped transition line of the transverse-field Ising model also persists, and its shape widens with decreasing $\chi$. At $\chi=\arccos(7/9)/2$ it turns into straight line, and bends further for even smaller values of $\chi$ until it merges with the boundary of the domain at $\chi=0$.

So while in the first scenario (Sec.\ \ref{s:1scenario}) the bow-shaped transition line of the transverse-field Ising model transforms into the zero-field transition line, both transition lines are simultaneously present in the second scenario (Sec.\ \ref{s:2scenario}). This difference already indicates that the first scenario cannot be mapped onto the second one by a simple rotation: Applying SO(3) rotation matrices to the spin components in the Hamiltonian \eqref{e:Hamiltonian} with $\lambda_1=1$ and $\lambda_2=\lambda_3=0$, cross-coupling terms proportional to $\sigma_k^1 \sigma_l^2$ emerge, and these appear to be responsible for the differences between the left- and right-hand columns in Fig.\ \ref{f:shu_general}.

\section{Equivalence of microcanonical and canonical ensembles}
\label{s:equivalence}

On the formal level, equivalence of the microcanonical and the canonical ensemble is related to the concavity of the microcanonical entropy in a straightforward way: Concavity is a necessary and sufficient condition for the ensembles to be equivalent. Notwithstanding, a naive study of concavity properties may lead to a misjudgment of whether observable differences exist between microcanonical and canonical results. A first example of this somewhat surprising statement was discussed in Sec.\ \ref{s:properties} of this paper: For the Curie-Weiss anisotropic quantum Heisenberg model, $e$ is uniquely determined by $\bm m$. For this reason, the same information can be encoded either in $s(\bm{m})$ or $s(e,m_1,m_2)$ and both these entropies describe the same physical situation of fixed energy $e$ and magnetization vector $\bm m$. Nonetheless, their concavity properties differ, as $s(\bm{m})$ is a concave function, while $s(e,m_1,m_2)$ is not. So which is the correct function whose concavity properties should be studied in order to draw conclusions about the (non)equivalence of ensembles? The answer depends on the choice of the ensemble one is comparing to. Typically this will be a canonical (or possibly a mixed canonical) ensemble where one of the thermodynamic variables is the inverse temperature $\beta=1/T$. Since $\beta$ is thermodynamically conjugate to the energy, it is the concavity properties of $s(e,m_1,m_2)$ [or $s(u,m_1,m_2)$, not discussed in this paper] that matter.

The entropy $s_{\bm{h}}(u)$, derived in Sec.\ \ref{s:shu_derivation} via a maximization procedure from the nonconcave entropy $s(e,m_1,m_2)$, is a concave function. This is in agreement with the fact that the Curie-Weiss anisotropic Heisenberg model \eqref{e:Hamiltonian} has a continuous temperature-driven phase transition, as the G\"artner-Ellis theorem \cite{Touchette11} excludes the possibility of a nonconcave entropy, and therefore of nonequivalent ensembles, in the absence of a discontinuous transition in the canonical ensemble. Despite this formal equivalence, the physical behavior of a thermally isolated (microcanonical) spin system can differ in an interesting way from its counterpart coupled to a heat bath. The reason for this is the fact that a conventional thermal bath consists of motional (or bosonic) degrees of freedom, and the inverse temperature $\beta$ of such a bath is positive. Microcanonically, positive inverse temperatures correspond to energies $u$ where
\begin{equation}
\beta=\frac{\partial s_{\bm{h}}(u)}{\partial u}>0.
\end{equation}
For all cases discussed in Sec.\ \ref{s:shu_general}, and independently of $\bm{h}$, this inequality is satisfied for all $u<0$. Therefore a system coupled to heat bath with positive $\beta$ can probe only those macrostates that correspond to negative energies. From the plots in the right two columns of Fig.\ \ref{f:shu_general} it can be seen that, for angles $\chi\in(0,\arccos(7/9)/2)$, the bow-shaped phase transition line is situated  in the region of positive energies, and is therefore inaccessible in a canonical setting with positive $\beta$. Microcanonically, by contrast, the entire range of energies $u$ in the domain of $s_{\bm{h}}(u)$ is accessible, and the phase transition can be probed in this ensemble. A similar situation of a phase transition line in the negative-temperature region of a spin system has been described for the two-dimensional Ising model and for the spherical model in \cite{KaPlei09,Kastner09}.

Both situations discussed in this section have concave entropy functions. While this implies some form of formal equivalence of ensembles, it does not necessarily imply that in practice (in the sense of the above discussion) microcanonical and canonical results will be identical, not even in the thermodynamic limit.

\section{Conclusions}
\label{s:conclusions}

In this article, we reported an exact, analytic computation of the microcanonical entropy $s_{\bm{h}}(u)$ of the anisotropic Curie-Weiss quantum Heisenberg model in the thermodynamic limit. The strategy of the calculation is to first obtain the entropy $s(\bm m)$ as a function of the magnetization vector $\bm m$. This is achieved by expressing, by means of the Trotter formula and a Hubbard-Stratonovich transformation, the microcanonical density of states $\Omega(e,\bm m)$ as a Laplace integral, and then evaluating this integral asymptotically in the large-system limit. The result \eqref{e:sem} for $s(\bm m)$ is remarkably simple and symmetric. The entropy $s_{\bm{h}}(u)$ as a function of the magnetic field vector $\bm h$ and the total energy per spin, $u=e-\bm h\cdot\bm m$, is obtained by maximizing $s(\bm m)$ under the constraint of fixed $\bm h$ and $u$.

On the basis of the plots in Figs.\ \ref{f:iso_isi} and \ref{f:shu_general} we discussed the properties of $s_{\bm{h}}(u)$, and in particular the phase transition lines along which the entropy is nonanalytic. We find two characteristic features: the zero-field transition line of the Ising model in a longitudinal magnetic field, separating ferromagnetic phases of different orientations; and the parabola-shaped transition line of the Ising model in a transverse magnetic field, separating an coupling-dominated phase inside the parabola from a field-dominated phase in the outside region. Varying the coupling constants and magnetic field components between these two extreme cases, we found coexistence of both types of transition lines, and coalescing or disappearing transition lines in other cases. For certain ranges of the coupling constants and fields, the phase transition line is situated in the region of negative absolute temperature. In a canonical setting with a heat bath restricted to positive temperatures, such a transition is unobservable, while it can be probed in a microcanonical setting at sufficiently large total energy $u$. 

Physically, the entropy $s_{\bm{h}}(u)$ describes a thermally isolated system with fixed energy $u$ in a magnetic field $\bm h$, but with fluctuating magnetization. This kind of study is motivated by recent experiments with cold atoms and ions that are isolated from their environment to an excellent degree, resulting effectively in a microcanonical setting. 
For more realistic long-range models with interactions decaying like $r^{-\alpha}$ with the distance $r$, the canonical free energy has been shown to coincide with the Curie-Weiss results in the thermodynamic limit for exponents $\alpha$ smaller than the lattice dimension. However, this does not hold for the microcanonical ensemble in the parameter region where microcanonical and canonical ensembles are nonequivalent \cite{Mori11,Mori12}.

\begin{acknowledgements}
We acknowledge useful comments by an anonymous referee who correctly pointed out to us that certain steps in derivation of the microcanonical entropy in Ref.\ \cite{KastnerJSTAT10} and also in the present paper are not rigorously justified.
 G.O.\ acknowledges financial support by the National Institute for Theoretical Physics, South Africa, and the Institute of Theoretical Physics, Stellenbosch University, South Africa, where he was based for the majority of this study. M.K.\ acknowledges support by the Incentive Funding for Rated Researchers program of the National Research Foundation of South Africa.
\end{acknowledgements}

\appendix

\section{Evaluation of \texorpdfstring{$\mathcal{F}$}{F} at the stationary points}
\label{s:evalF}
It is shown how to evaluate $\mathcal{F}$ as given in \eqref{e:F2} at a stationary point determined by equations \eqref{e:saddle3a}--\eqref{e:saddle3c}.

We start by writing \eqref{e:saddle3b} in the form
\begin{equation}\label{e:name}
-\frac{m_\alpha r}{\tanh r} = t_\alpha-x_\alpha\sqrt{\lambda_\alpha}.
\end{equation}
Upon squaring, summing over $\alpha$, and then taking the square root on both sides of this equation, we obtain
\begin{equation}\label{e:r_selfcon}
\frac{|\bm{m}|r}{\tanh r}=\sqrt{\sum_{\alpha=1}^3\left(t_\alpha-x_\alpha\sqrt{\lambda_\alpha}\right)^2}=r,
\end{equation}
where the second equality sign is due to definition \eqref{e:Rt1}. We therefore have $\tanh r=|\bm{m}|$ and can write
\begin{equation}\label{e:Fintermed}
\mathcal{F}(s,\bm{t},\bm{x})=2es+\bm{m}\cdot\bm{t}-\frac{1}{2}\ln\left(1-\bm{m}^2\right).
\end{equation}
The first two terms on the right hand side of \eqref{e:Fintermed} can be written in the form
\begin{equation}
\bm{m}\cdot\bm{t}+2es
=\sum_{\alpha=1}^3\left(m_\alpha t_\alpha - \frac{1}{s}x_\alpha x_\alpha\right)
=\sum_{\alpha=1}^3 m_\alpha\left(t_\alpha - x_\alpha \sqrt{\lambda_\alpha}\right),
\end{equation}
where first \eqref{e:saddle3a} and then \eqref{e:saddle3c} have been used. With \eqref{e:name} and \eqref{e:r_selfcon}, this expression simplifies to
\begin{equation}
\bm{m}\cdot\bm{t}+2es
=-\frac{\bm{m}^2 r}{\tanh r} = -|\bm{m}|r = -|\bm{m}|\arctanh|\bm{m}|.
\end{equation}
Inserting this into \eqref{e:Fintermed}, the derivation of \eqref{e:Ffinal} is complete.


\end{document}